\documentclass[aps,pra,preprint,groupedaddress, showpacs, 11pt] {revtex4} 
\usepackage{amsmath} 
\usepackage[pdftex]{graphicx} 
\usepackage{hyperref}
\usepackage{multirow}
\usepackage{hhline}

\newcommand{\vect}[1]{\boldsymbol{#1}}

\begin{document}

\title{Angle-Resolved Spectroscopy of Parametric Fluorescence}
\author{Feng-Kuo Hsu}
\author{*Chih-Wei Lai}
\affiliation{Department of Physics and Astronomy, Michigan State University, East Lansing, MI 48824}
\email{cwlai@msu.edu} 

\date{\today}

\begin{abstract}
The parametric fluorescence from a nonlinear crystal forms a conical radiation pattern. We measure the angular and spectral distributions of parametric fluorescence in a beta-barium borate crystal pumped by a 405-nm diode laser employing angle-resolved imaging spectroscopy. The experimental angle-resolved spectra and the generation efficiency of parametric down conversion are compared with a plane-wave theoretical analysis. The parametric fluorescence is used as a broadband light source for the calibration of the instrument spectral response function in the wavelength range from 450 to 1000 nm.  
\end{abstract}

\pagenumbering{roman}
\maketitle

\pagenumbering{arabic}
\section{Introduction}
Signal--idler photon pairs generated by spontaneous parametric down-conversion (SPDC) have been used to address fundamental issues of quantum theory and have found application in quantum entanglement and quantum information processing and metrology \cite{Shih1988,Shih1994,Kwiat1995, Shih2003}. The SPDC process, also known as parametric fluorescence or parametric scattering \cite{Louisell1961, Gordon1963, Kleinman1968, Harris1967, Giallorenzi1968, Byer1968}, is a second-order optical process in which a driving pump photon is scattered into signal--idler photon pairs subject to energy and momentum conservation. This spontaneous parametric emission can be described properly only by field quantization. Since its prediction and observation in the 1960s, parametric fluorescence has become a technique for measuring second-order nonlinear susceptibilities \cite{Byer1968, Choy1976, Shoji1997, Shoji2002} and for developing tunable light sources via parametric oscillation or amplification processes. In 1969, Zeldovich and Klyshko \cite{Zeldovich1969} first proposed the use of parametric fluorescence (luminescence) as a nonclassical source of photon pairs. This description was experimentally verified by Burnham et al. in 1970 \cite{Burnham1970}.

The possible wave vectors of the signal--idler photon pairs are determined by energy and momentum conservation, a constraint referred to as phase-matching, leading to highly directional parametric emission. The phase-matching condition frequently cannot be met for specific wavelengths of interest or practical applications owing to the limited tunability of inherent dispersion of nonlinear materials. However, it can be met by selecting polarization birefringent crystals with appropriate refractive indices or by designing waveguides or periodic structures of specific wavelengths. There are two major types of polarization phase-matching schemes for parametric down-conversions: Type-I, where signal--idler photons have the same polarization (co-linearly polarized photons), and Type-II, where the signal--idler photons have orthogonal polarization (cross-linearly polarized photons). Both types of parametric processes have been used to generate photon pairs, sometimes referred to as biphoton states, which exhibit correlation/entanglement for variables including polarization, momentum, time, energy, and angular momentum. 

When the phase-matching condition is met, the signal and idler radiation form a conical pattern independent of the intensity of the pump source. The angular distribution of parametric fluorescence is determined by the energy of the pump, signal, and idler waves, subject to the dispersion of the crystal and walk-off angles of these three waves. The magnitude of the second-order nonlinear susceptibility $\chi^{(2)}$ is a typical selection criterion for parametric downconversion. Many uniaxial or biaxial nonlinear crystals have been used for parametric down conversion: for example, KD*P (potassium dideuterium phosphate, $\rm{KD_2PO_4}$), BBO (beta-barium borate, $\rm{\beta-BaB_2O_4}$), and LBO (lithium niobate, $\rm{LiNbO_3}$). In this report, we use BBO, a negative uniaxial class 3m crystal characterized by a wide range of transparency over the ultraviolet ($\lambda\approx 200$nm) to the infrared ($\lambda\approx 3500$nm) portion of the spectrum  \cite{Chen1985}. BBO crystals have been widely studied for harmonic frequency generation, optical parametric oscillation, and generation of bi-photon states. 

\section{Experimental Methods}
We measure the angular distribution and photon flux of parametric fluorescence from a 3-mm thick BBO crystal. The BBO crystal is cut at an angle of $\theta_{m} = 29 \pm0.5 \, ^{\circ}$ with respect to the optical axis. This cut angle $\theta_{m}$ is chosen for the Type-I ($e \rightarrow o \, + o$) degenerate parametric down-conversion at $\lambda= 810$ nm with a pump $\lambda _p = 405$ nm. The crystal is mounted on a three-axis rotary mount with the crystal's optical axis (OA) in the horizontal plane. The angle formed by the OA and the pump's propagation wave vector can be finely tuned by tilting the crystal to satisfy the phase-matching condition for various $\theta_m$ near the crystal cut angle. We can thus adjust the pump and signal Poynting vectors from collinear to non-collinear and generate parametric fluorescence with varying conical emission angles. 

The pump is a violet diode laser with a $TEM_{00}$ linearly polarized 2-mm $1/e^{2}$ diameter output beam at a wavelength $\lambda _p = 405 \, nm$ (CNI Laser MLL-III-405). The pump beam is focused on the crystal through a lens (L1) with a focal length of $500 \, mm$. Lens L1 and the objective are positioned to form a telescope such that the residual pump beam is collimated with a reduced beam radius below $100 \mu m$. By passing the pump beam through a pair of a half-wave plate (HWP) and a Glan--Taylor polarizer (P1), we can vary the incident pump intensity by rotating the HWP while maintaining the degree of linear polarization better than $99.9\%$. 

\begin{figure}[!tbp]
\includegraphics[width=0.75 \columnwidth]{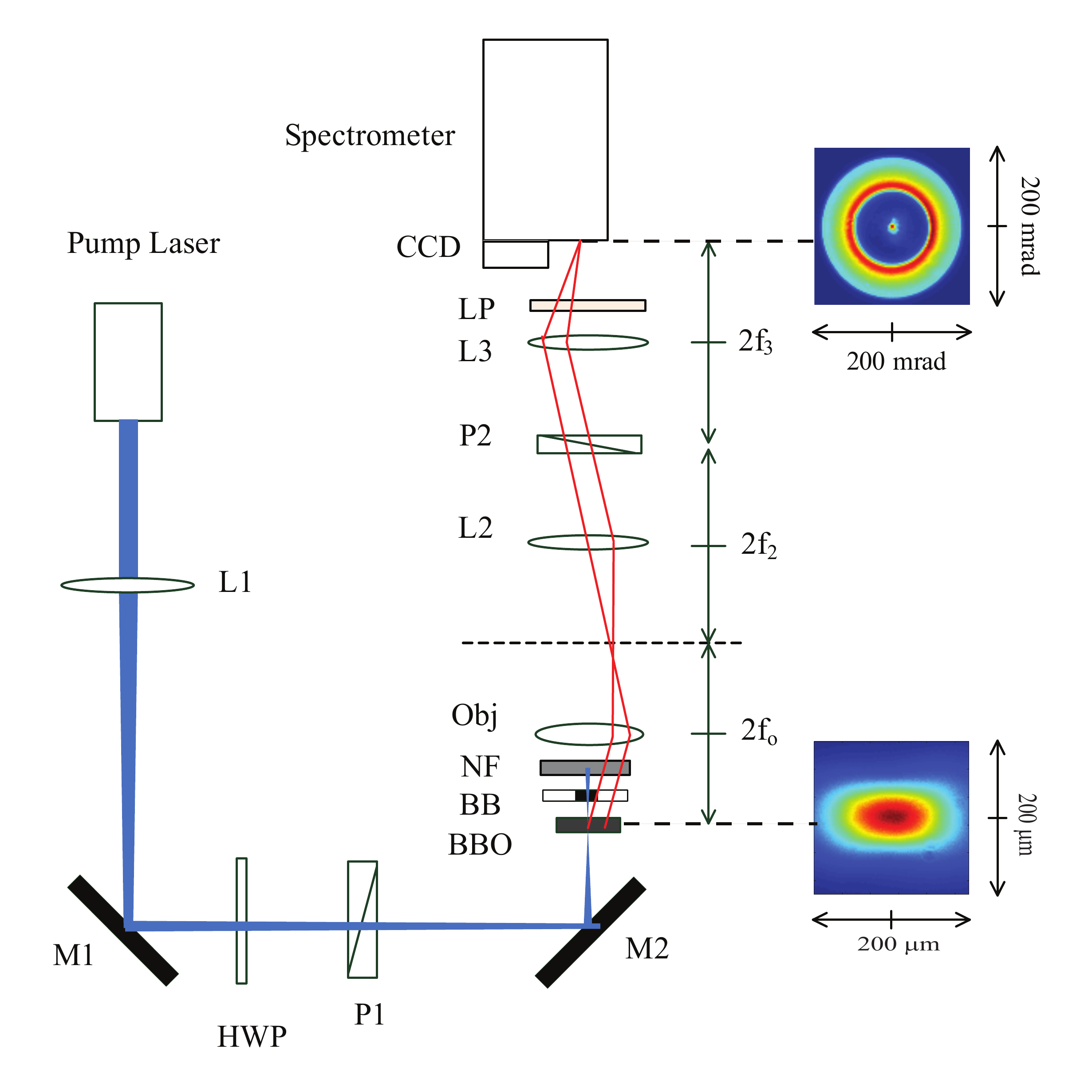}
\caption{{\bf Schematic diagram of the experimental setup.} The transmitted and scattered pump photons are rejected by a miniature beam blocker, a thin-film notch filter, and long-pass filters. The angular and spectral distributions of conical parametric fluorescence are measured by an imaging spectrometer through a Fourier transform optical system. L1, L3 and L4 are convergent lenses with focal length f = 500, 100, and 100 mm, respectively; L2 is an objective ($20 \times$ N.A.=0.26) with an effective focal length of 10 mm (Mitutoyo Plan Apo infinity-corrected long-working-distance objective); P1 and P2 are Glan--Taylor and Glan--Thompson polarizers; M1 and M2 are silver mirrors; HWP is a half-wave plate for $\lambda=405$ nm; BB is a miniature pump beam blocker; NF is a notch filter (Semorck 405-nm StopLine single-notch filter); and LP represents longpass filters (a Semrock 409-nm blocking edge BrightLine® long-pass filter and a Schott GG435 glass filter). When L2 is removed, a real-space fluorescence image is formed at the entrance of the spectrometer with an imaging magnification of $10\times$. Examples of real-space and angle-resolved fluorescence images are shown with actual dimensions. \label{fig:PFExpSetup}}
\end{figure}

The angle-resolved images (Fig. \ref{fig:ARImages}) and spectra (Fig. \ref{fig:ARSpectra}) of parametric fluorescence are measured by a Fourier transform optical system, including a $20 \times$ long-working-distance objective and an imaging spectrometer as shown in Fig. \ref{fig:PFExpSetup}. The BBO crystal is positioned at the focal plane of the objective lens (effective focal length $f_{o} = 10 \, mm$). The parametric fluorescence with an amplitude distribution $F(x,y)$ at the crystal is collected by a $20 \times$ microscope objective with a 10-mm effective focal length (numerical aperture N.A. = 0.26). The back focal plane of the objective is the Fourier transform plane with coordinates $(u,v)=(f_{o}\times sin(\theta_{x}), f_{o}\times sin(\theta_{y}))$. The collection angle is within $\pm 15^\circ$, limited by the objective. The objective lens converges parallel rays emanating from the crystal to the back focal plane of the objective. In this plane, the fluorescence image in the crystal is transformed into a far-field image in spatial frequency that is related to the emission angle as described above. The spatial intensity distribution of parametric fluorescence in the back focal plane of the objective lens thus corresponds to the angular distribution of radiation. This Fourier transform plane is placed at the front focal plane of lens L2 (focal length $f=100 \, mm$). Lenses L2 and L3 are identical and separated by a distance of 2f, and they relay the Fourier transformed images to the entrance plane and then onto the charge-couple device (CCD) through the zero-order diffraction off the grating of the imaging spectrometer (PI-Acton SpectroPro 2750i, focal length 750 mm). In this way, we measure the angular distribution of parametric fluorescence. When lens L2 is removed, we project the real-space spatial intensity distribution of parametric fluorescence from the BBO crystal onto the CCD.  

The fluorescence image is recorded by a CCD positioned in a conjugate imaging plane of the Fourier plane. The resultant intensity distribution is related to the Fourier transform of the intensity of the parametric fluorescence $I(x,y)=|F(x,y)|^2$. By projecting this far-field image through the entrance slit and the first-order diffraction of a 300 lines/mm grating, we obtain angle-resolved spectra as an image by taking the spectral dispersion of the parametric fluorescence as a function of angle. The spectral resolution of 0.1 nm is determined by the dispersion of the grating and the width of the entrance slit ($\approx 100 \mu$m). The spatial and angular resolutions of the system are approximately 2 $\mu m$ and 2 mrad, respectively, limited by the pixel size of the liquid-nitrogen-cooled CCD camera. 

In a parametric scattering process, only about one out of $10^{10}$ incident photons is parametrically down-converted. It is essential to prevent the transmitted and scattered pump photons from entering the spectrometer. For Type-I phase matching in a negative uniaxial BBO crystal (e-o-o case), the polarization of the high-frequency pump (e-wave) is orthogonal to the polarization of the signal and idler (o-waves). Thus, the transmitted and scattered pump photons can be suppressed by approximately 4--5 orders of magnitude by a pair of polarizers (P1 and P2) with orthogonal polarization orientations for pump and single/idler waves, respectively. The pump photons are further rejected by a thin-film notch filter (Semrock 405-nm StopLine single-notch filter) and two longpass filters (a Semrock 409-nm blocking edge BrightLine® long-pass filter and a Schott GG435 glass filter). The filters are arranged in the sequence shown in Fig. \ref{fig:PFExpSetup} to suppress fluorescence from filters induced by the transmitted violet pump laser. The combination of polarizers and filters allows for a rejection of pump photons by approximately a factor of $10^{10}$. This rejection ratio of $10^{10}$ can be further improved above $10^{14}$ by a miniature beam blocker made of a $\sim 0.5$ mm-diameter silver-paste dot on a microscope cover positioned in front of the notch filter (NF)/objective. Depending on the signal wavelength, the measured angle-resolved spectra may still contain residual transmitted and scattered pump and background fluorescence from filters. We measure such a background emission spectrum in the absence of Type-I e-o-o parametric fluorescence by rotating the BBO crystal 90 degrees azimuthally.

\section{Experimental Results and Modeling}
\subsection{Phase Matching}
The three-wave parametric processes are calculated according to the conservation of energy and momentum, commonly referred to as phase matching. The angle-resolved spectra of the parametric fluorescence are consistent with the tuning curves calculated for the phase-matching condition under a plane-wave approximation. 

The energy conservation condition is expressed as 
\begin{equation}
\omega_{p} = \omega_{s}+ \omega_{i} \, , \label{eq:energy}
\end{equation}
where $\omega_{p}$ is the frequency of the incident pump wave and $\omega_{s}$ and $\omega_{i}$ are the frequencies of the signal and idler waves. 

The momentum conservation condition can be expressed as
\begin{equation}
\textbf{k}_{p} = \textbf{k}_{s}+ \textbf{k}_{i} \, , \label{eq:momentum}
\end{equation}
where $\textbf{k}_{p}$, $\textbf{k}_{s}$, and $\textbf{k}_{i}$ are the pump, signal, and idler wave vectors, respectively. For Type-I down-conversion in a BBO, the signal and idler labels are arbitrary. In the case of degenerate down conversion, $k_{s}  = k_{i}$, and Eq. (\ref{eq:momentum}) reduces to
\begin{equation}
n_{p} = n_{s} \\ cos (\theta'_{s}) \, , \label{eq:index}
\end{equation} 
where $n_{p}$ and $n_{s}$ are the indices of refraction of the pump and signal, and  $\theta'_{s}$ is the angle formed by the propagation directions of the signal and pump waves inside the crystal.

\begin{figure}[!tbp]
\includegraphics[width=1.0 \columnwidth]{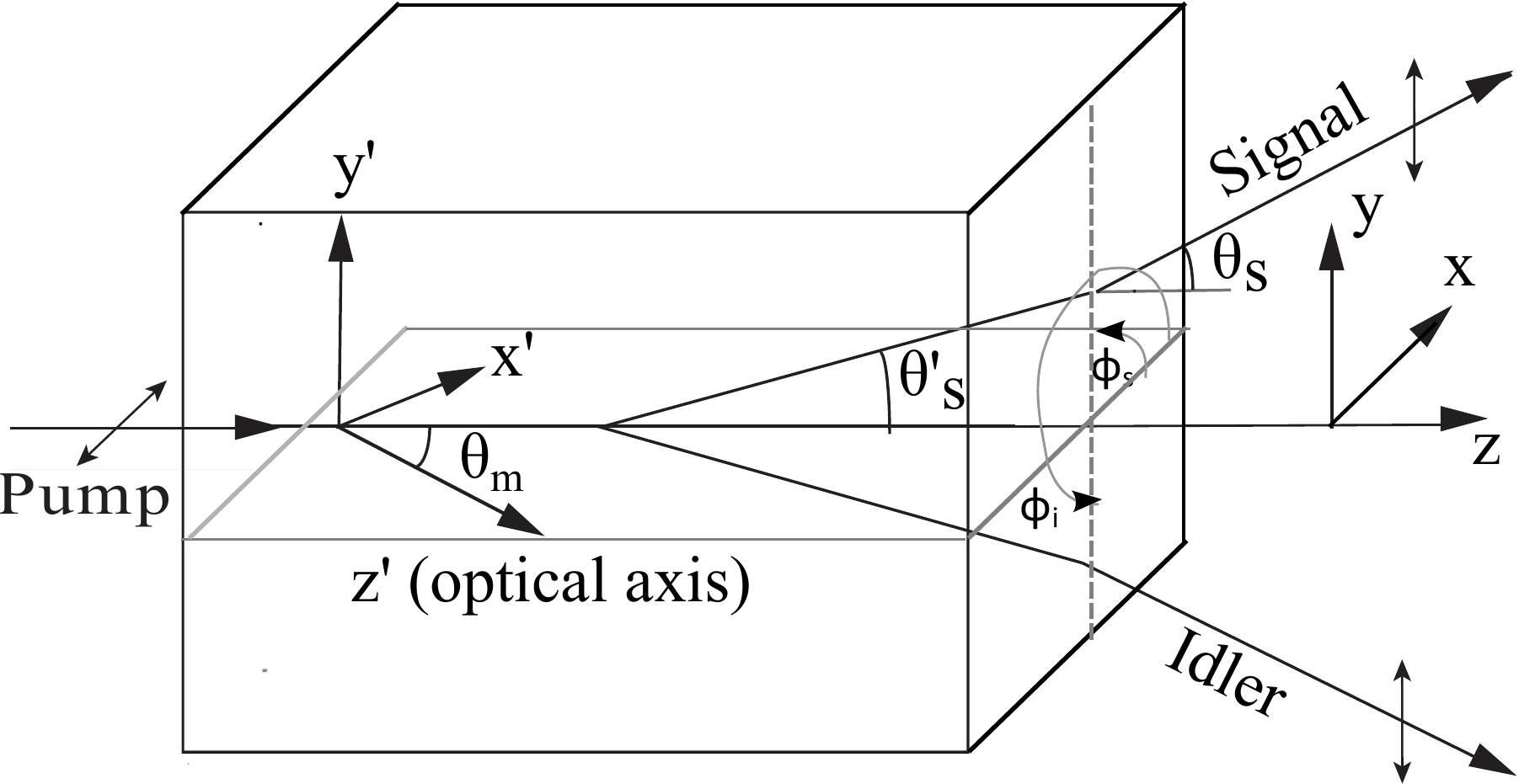}
\caption{ {\bf Schematic diagram of the crystal and laboratory frame coordinates for parametric down-conversion in a BBO crystal.}  The phase-matching angle $\theta_{m}$ is defined as the angle formed by the crystal optical axis ($z'$) and the pump wave vector ($z$). The angles $\theta' _{s}$ and $\theta _{s}$ are, respectively, internal and external angles formed by the signal and pump wave vectors. Here the incident pump wave is horizontally polarized, leading to a vertically polarized down-converted signal and idler waves. \label{fig:BBOAxes}} 
\end{figure}

\begin{figure}[!tbp]
\includegraphics[width=1.0 \columnwidth]{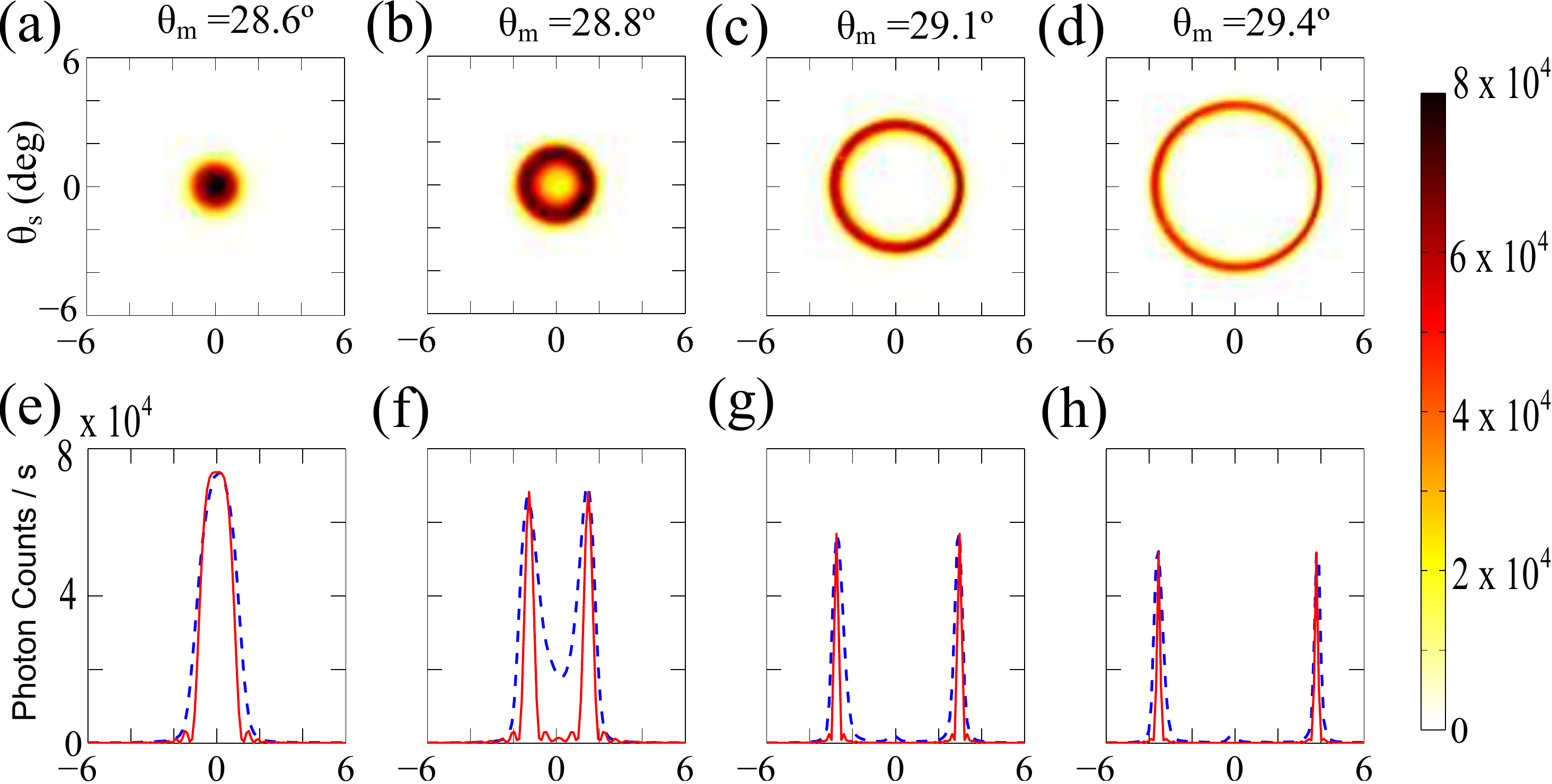}
\caption{{\bf Angular intensity distribution of parametric fluorescence at 810 nm.} (a)--(d) Angle-resolved images of parametric fluorescence for $\lambda = 810\pm0.5$ nm and $\theta_{m}$ as indicated. The external angle $\theta_s$ is formed by the pump and signal wave vectors in air (see also Fig. \ref{fig:BBOAxes}). The color palette represents the intensity of the single radiation. Parametric fluorescence is spectrally filtered through a 1-nm bandpass filter with central wavelength $\lambda=800$ nm. The phase-matching angle is adjusted by tilting the BBO crystal with respect to the pump wave vector. The collinear phase-matching angle is set to $\theta_{m}=28.6^\circ$ according to a theoretical calculation using indices of refraction given in Ref. \cite{Boeuf2000}. (e)--(g) Experimental (blue dashed line) and theoretical (red solid line) cross-sections. The non-collinear phase-matching angle relative to the collinear one can be determined experimentally by measuring the tilting angle of the crystal surface with respect to the pump wave vector. \label{fig:ARImages}}
\end{figure}

We use a BBO crystal cut at an angle of $\theta_{m} = 29 \pm0.5 \, ^{\circ}$, optimized for Type-I parametric down-conversion ($e \rightarrow o + o$). Down-converted signal/idler photons are co-linearly polarized but orthogonal to the polarization of the pump wave. The wavelengths and wave vectors of the parametric fluorescence are determined by the phase-matching angle $\theta_{m}$, the angle formed by the optical axis of the crystal (z'-axis), and the wave vector of the pump wave (z-axis) as shown in Fig. \ref{fig:BBOAxes}. The crystals are mounted on a rotation stage such that the optical axis lies in a horizontal plane when the parametric fluorescence signal is maximized. By tilting the crystal, we can vary the phase-matching condition from collinear to non-collinear, leading to a conical angle up to 5 degrees for degenerate parametric fluorescence near 810 nm. 

For Type-I phase matching, the incident pump photons are subject to the extraordinary index of refraction $\tilde{n}(\theta_{m})$, while the down-conversion photons are subject to the ordinary index of refraction. The extraordinary index of refraction, $\tilde{n}$, depends on the phase-matching angle $\theta_{m}$ and follows the relationship:
\begin{equation}
\tilde{n}(\theta_{m},\lambda) = \left(\frac{\cos(\theta_{m})^2}{n_{o}(\lambda)^2} + \frac{\sin(\theta_{m})^2}{n_{e}(\lambda)^2}\right)^{-\frac{1}{2}}\, . \label{eq:npe2}
\end{equation}

In the parametric process, a pump wave of wavelength $\lambda_{p}$ creates signal waves at $\lambda_{s}$, and angles $\theta_{s}$, subject to energy conservation (Eq. (\ref{eq:energy})) and momentum conservation (Eq. (\ref{eq:momentum})). In the Type-I e-o-o case, $n_{s}=n_{o}(\lambda_{s})$ and $n_{p}=\tilde{n}(\theta_{m},\lambda_{p})$ (Eq. \ref{eq:npe2}). Here the labeling of signal and idler waves is arbitrary ($\theta_s=\theta_i$). A continuum of phase-matching functions  $\Phi(\lambda_{s},\theta_{s})$ for parametric fluorescence can be obtained using the aforementioned equations and indices of refraction $n_{o}(\lambda_s)$ and $n_{e}(\lambda_s)$.  Indices of refraction of wavelengths ranging from 0.3 $\mu$m to 5 $\mu$m are extracted from "NIST Noncollinear Phase Matching in Uniaxial and Biaxial Crystals Program" as described in Ref. \cite{Boeuf2000}. We calculate the phase matching functions (tuning curves) for down-converted signal/idler waves ranging from 430 to 1000 nm for a pump wave $\lambda_p=405$ nm. 

\subsection{Angle-Resolved Imaging}

We adopt the plane-wave analysis developed in Refs. \cite{Koch1995, Kleinman1968, Hong1985} to determine the angular distribution of parametric fluorescence. The effects of a finite pump beam size have also been considered in, for example, Refs. \cite{Ling2008, Mitchell2009}. Under a plane-wave approximation, the parametric fluorescence forms a conical angular distribution The diameter and axis of the conical emission are determined by the wavelengths of the pump, signal, and idler waves, the Poynting vector walk-off angles of the three interacting waves, and the dispersion of the nonlinear crystal. In such a parametric process, the angular spread of the conical emission is determined by the conservation of transverse and longitudinal momenta of interaction waves. The transverse momentum induced by focusing the pump into the crystal contributes to a finite angular spread of the cones. By considering the longitudinal momentum, we deduce that parametric fluorescence signal is inversely proportional to the interaction length. In our experiments, we use a lens with a long focal length ($f=500$ mm), resulting in a focal spot with a $1/e^2$ radius larger than $100 \mu$m. Thus, for the experiments reported here, the effects of walk-off and finite pump beam size are negligible compared with the spectral and angular resolution of the optical system. 

We can determine the angular distribution of parametric fluorescence using the following phase-matching function for a finite crystal length $L$ and a pump Gaussian beam profile with a $1/e^2$ radius $W$ \cite{Rubin1996, Boeuf2000, Fedorov2008}:

\begin{align} 
\Phi &= \exp \left( { - \frac{1}{2}{W^2}\left( {\Delta k_x^2 + \Delta k_y^2} \right)} \right) \cdot {\left( {\frac{{\sin \left( {\frac{1}{2}L\Delta {k_z}} \right)}}{{\frac{1}{2}L\Delta {k_z}}}} \right)^2} \nonumber \\
&= \exp \left( { - \frac{1}{2} \kappa^2 {W^2}} \right) \cdot {\rm{sinc}^2}{\left( {\frac{1}{2}L\Delta {k_z}} \right) } \, . \label{eq:Phi}
\end{align}
 
The mismatch wave vector, $\Delta \vect{k}$, is decomposed into longitudinal ($\hat{z} \parallel \vect{k}_p$) and transverse (in x-y plane) parts: $\Delta k_z$  and $\kappa=\sqrt{\Delta k_x+\Delta k_y}$. The phase-matching tolerances can be considered in terms of the angular spread ($\Delta \theta_s$) and spectral bandwidth ($\Delta \lambda_s$), defined as the full-width-at-half-maximum (FWHM) for the above function. For a pump wave with a focal radius $W \approx 50 \mu$m, only the tolerances from the ${\rm{sinc^2}}$ part can be measured in our optical system. For this situation, considering a Taylor series expansion of $\Delta k_z$ near the perfect phase-matching point ($\Delta k_z=0$), we can analytically deduce the angular spread ($\Delta \theta_{\rm{FWHM}}$) and spectral bandwidth ($\Delta \lambda_{\rm{FWHM}}$) for the degenerate case ($k_s=k_i$): 

\begin{equation}
\Delta\theta_{\rm{FWHM}}(\theta_s)= \frac{2 \times 0.886\,\pi}{L \times |\partial \Delta k_z/\partial \theta_s|} \approx \frac{2.783 \times n_{s} }{L \times k_s \times \theta_s }\,, \label{eq:angular_spread}
\end{equation}
and 
\begin{equation} 
\Delta\lambda_{\rm{FWHM}}(\lambda_s)= \frac{2 \times 0.886\,\pi}{L \times |\partial \Delta k_z/\partial \lambda_s|} \approx \frac{ 0.443\times  {\lambda_s}^2}{L \times n_{s} \cos(\theta_s/n_{s}) }\, \nonumber,
\end{equation}
where $\Delta k_z = |{k_p}' - 2 \, {k_s}' \cos ({\theta_s}')|$, ${k_{s}}' = n_{s} \omega_{s}/c =  2 \pi \, n_{s} / \lambda_s$, and $\theta_s \ll 1$.

\begin{figure}[!tbp]
\includegraphics[width=1.0 \columnwidth]{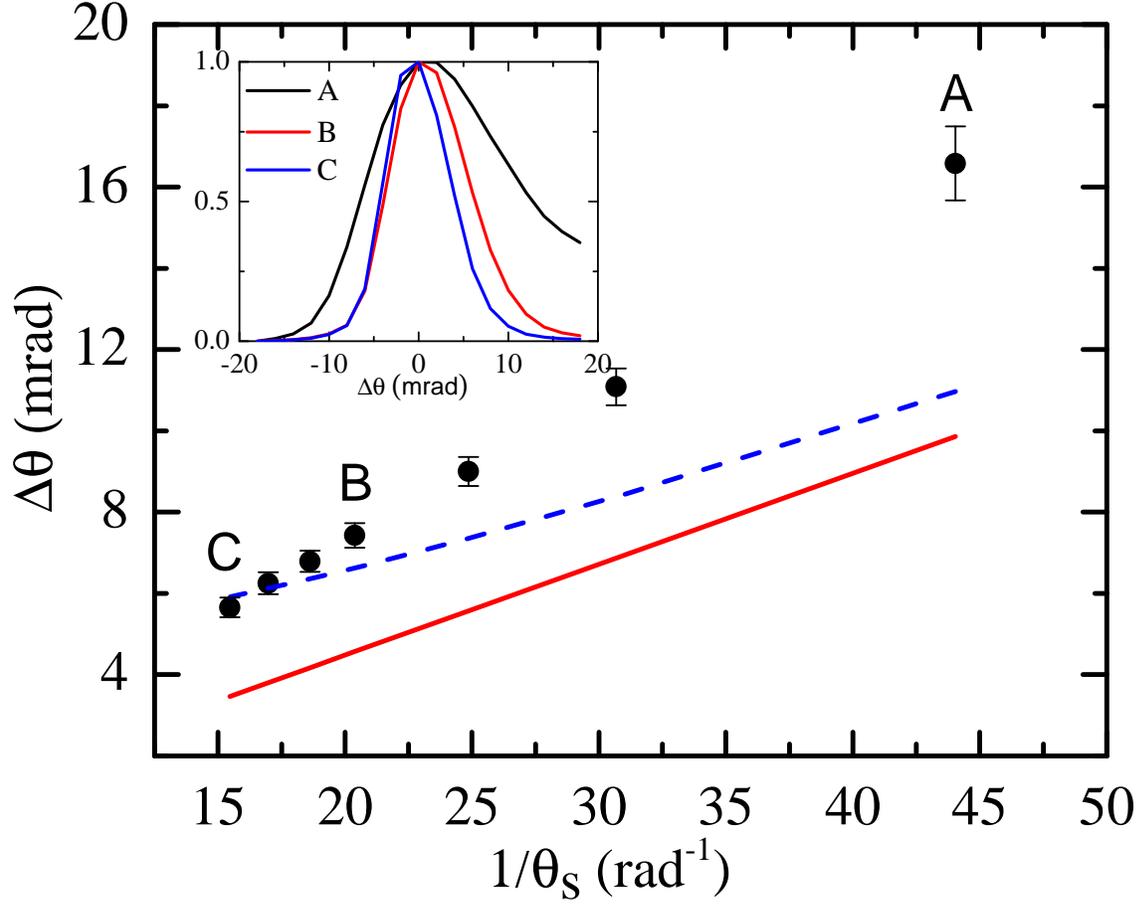}
\caption{{\bf Angular spreads $\Delta \theta_{\rm{FWHM}}$ of parametric fluorescence at 810 nm.} $\Delta \theta_{\rm{FWHM}}$ is plotted as a function of the inverse of the signal angle ($1/\theta_{s}$). Experimental data are represented with error bars as solid circles. The red solid line is the theoretical curve according to Eq. \ref{eq:angular_spread}, while the dashed line is the theoretical curve including a finite angular resolution of 2 mrad. Selected  experimental angular intensity profiles for A ($\theta_s=1.3^\circ=0.023$ rad), B ($2.8^\circ=0.049$ rad), and C ($3.7^\circ=0.065$ rad) are shown in the inset.\label{fig:AngularSpread}}
\end{figure}

The parametric fluorescence signal of angle-resolved images at $\lambda=810$ nm are shown in Fig.~\ref{fig:ARImages}. These false-color images, taken through a 1-nm band-pass filter, represent the angular intensity distributions of parametric fluorescence at $\lambda=810 \pm 0.5$ nm for the phase-matching angle $\theta_m=28.6^{\circ}, \, 28.8^{\circ}, \, 29.1^{\circ}$, and $29.4^{\circ}$. The BBO crystal is cut at the designed angle with about $1^\circ$ tolerance. To determine the phase-matching angle $\theta_m$ with better precision, we first set the phase-matching angle for the collinear case by comparing the simulated and experimental angular distributions. We then deduce the phase-matching angle for the non-collinear case from the tilting angle of the crystal relative to that for the collinear case. The conical signal angle ($\theta_s$) of degenerate parametric fluorescence at $\lambda=810$ nm increases with $\theta_{m}$. In Fig. \ref{fig:AngularSpread}, we plot the angular spread ($\Delta \theta_{\rm{FWMH}}$) as a function of the inverse of the signal angle ($1/\theta_s$). The angular spread decreases with $\theta_{s}$ for small angles when $\sin(\theta_s) \approx \theta_s$. We attribute the discrepancy between experimental data and Eq. (\ref{eq:angular_spread}) to a limited experimental angular resolution ($\approx 2$ mrad), finite pump beam size and spatial coherence, and the birefrigent walk-off.  

\subsection{Angle-Resolved Spectroscopy}

The parametric fluorescence flux per unit frequency is 
\begin{widetext}
\begin{align}
N_s(\omega _s, \vect{\kappa}_s) = \frac{\hbar \, {d_{eff}}^2 \, {\omega _s} \, {\omega _i} \, {\omega_p} \, L^2 \, N_p }{2 \pi^4 \, c^3 \, \epsilon _0 \, n_s \, n_i \, n_p} \, d\omega _s \, d^2 \vect{\kappa}_s \int d^2 \vect{\xi} \, \exp (\frac{1}{2} {\xi}^2) \cdot \rm{sinc}^2 \left( {\frac{1}{2}\,L\,\Delta {k_z}} \right), \label{eq:photon_flux}
\end{align}
\end{widetext}
where $d_{eff}$ is the effective second-order nonlinear coefficient, L the interacting crystal length, $N_p=P_p/\hbar \omega_p$ the pump flux, $\vect{\xi}=\vect{\kappa} W= (\vect{\kappa}_s +\, \vect{\kappa}_i) \,W$ the dimensionless transverse momentum associated with the signal and idler waves, and W the $1/e^2$ pump beam radius. The integration over $\vect{\xi}$ is detailed in Appendix.

Experimental angle resolved spectra $N^*_s(\lambda_s, \theta_s)$ and corresponding calculated $N_s(\lambda_{s}, \theta_{s})$ are shown in Fig.~\ref{fig:ARSpectra} (e)--(h). The tuning curves, corresponding to the perfect phase-matching condition, are shown as white dashed lines on the experimental spectra. The experimental parameters are $N_p=1.63 \times 10^{17} $ /s ($P_p=80$ mW ), L=3 mm, and $W^2= 60 \mu m \times 30 \mu m$. The indices of refraction $n_s$, $n_i$, and $n_p$ are evaluated using the database in Ref. \cite{Boeuf2000} ($n_{s,i} \approx 1.66$ at 810 nm). The effective nonlinear coefficient of a BBO crystal can be deduced from its d-matrix using $d_{eff}=d_{31} \, \sin(\theta_m+\rho)-d_{22}\cos(\theta_m+\rho) \, \sin(3 \phi)$ for Type-I phase matching. $\theta_m$ is the phase-matching angle, $\phi$ the azimuthal angle, and $\rho$ the birefringent walk-off angle. We use $d_{eff}$=1.75 pm/V for 405 nm adopting from Refs. \cite{Shoji2002,Shih2003,Klein2003}. 

Computationally, the angular distributions of parametric fluorescence are calculated for signal wavelengths from 430 to 1000 nm with step size $\delta \lambda =1$ nm and $\delta \theta _s = 0.03^\circ \approx 0.5$ mrad. Selected calculated angular distributions of fluorescence flux per 1 nm are shown in Fig.~\ref{fig:ARSpectra} (e)--(h) for $\theta_m=28.6^\circ, \, 28.8^\circ, \, 29.1^\circ, \, 29.4^\circ$. The shortest signal wavelength appears in the simulation is $\approx 440 \, nm$, limited by the availability of refractive indices between $\lambda=0.3 \, \mu m$ and $5 \, \mu m$ \cite{Migdall2000}. Experimentally, parametric fluorescence with a wavelength as short as $431 \, nm$ can be observed near $\theta_s=0$. 

The experimental angle-resolved parametric fluorescence images are shown in the left panel of Fig.~\ref{fig:ARSpectra}. These images are taken through a notch filter and longpass filters (cutoff wavelength of 420 nm) as shown in Fig.~\ref{fig:PFExpSetup}. Experimental angle-resolved spectra are acquired by spectrally resolving the parametric fluorescence across the fluorescence cone center through the entrance slit with an opening of 100 $\mu$m, corresponding to $\Delta \theta_x \approx 0.6^\circ$. Angle-resolved fluorescence images (Fig.~\ref{fig:ARSpectra}(a)--(d), left panel) are taken for the zero-order diffraction of a holographic grating with 1800 lines/mm, while angle-resolved spectra (right panel) are dispersed by a  grating with 300 lines/mm and a blaze wavelength of 1000 nm. The angular resolution is about 2 mrad, while the spectral resolution is about 0.1 nm. Residual stray or scattered pump laser light can be measured by rotating the BBO crystal $90^{\circ}$ azimuthally. Such background 'noise' is subtracted. Note that the measured CCD intensity is subject to the nonuniform spectral sensitivity and collection efficiency of the optical system including the effects of lens coating, optical filters, and gratings, and the spectral response of the CCD camera. The spectra branches out from $\theta_{s}=0$ at $\lambda \approx 433 \, nm$. The secondary weak arcs in the wavelength range above 870 nm outside of the expected parametric fluorescence peaks are due to the second-order diffraction of the grating for the fluorescence from approximately 435 nm to 500 nm. The inner arcs branching from 435 nm and closing near 530 nm are due to parametric fluorescence from a Type-II parametric process ($e \rightarrow o\,+e$). The turning curves for such Type-II parametric fluorescence are indicated by black dashed-doted lines in experimental spectral images. 

\subsection{Fluorescence Photon Flux}
The integrated parametric fluorescence photon flux can be obtained by the integrating over $\vect{\xi}$ and $\vect{\kappa}_s$ in Eq.~(\ref{eq:photon_flux}). For values of $\theta_m$ or $\omega_s$ such that the parametric fluorescence cone has a radius sufficiently large with negligible emission at the cone center, the resultant integrated fluorescence flux is \cite{Koch1995}

\begin{align}
N_s &= \frac{\hbar \, {d_{eff}}^2 \, L \, {\omega_s}^2 {\omega_i}^2} {\pi \, c^4 \, \epsilon_0 \, {n_p}^2}   \, {N_p} \, d \omega_s \nonumber \\ 
&= (2 \pi)^4 \, \frac{2 \hbar \, c \, {d_{eff}}^2 \, L}{\epsilon_0 \, {n_p}^2 \, {\lambda_s}^4 \, {\lambda_i}^2} \, N_p \, d \lambda_s \, .  \label{eq:int_flux}
\end{align}
The efficiency $\eta_s \equiv N_s/N_p$ is a coefficient depending largely on the material properties such as the second-order nonlinear coefficient, interacting crystal length, and index of refraction for the pump wave. Assuming a bandwidth $\Delta \lambda=1$ nm and $L=3$ mm, the efficiency coefficient $\eta_s = 1.3 \times 10^{-10}$ for the degenerate parametric fluorescence at $\lambda_s=810$ nm under $\lambda_p=405$ nm. Specifically, we evaluate $\eta_s$ for $\theta _m = 29.12^{\circ}$ (Fig.~\ref{fig:ARSpectra}g). We integrate the photon flux of simulated angle-resolved spectra for $\lambda _s = 809.5 \rightarrow 810.5$ nm, $\theta _s = 0 \rightarrow 7.5^\circ$, and $\phi=0 \rightarrow 2 \pi$. The total parametric fluorescence photon flux, including both degenerate signal and idler waves, is $2 \, N_s \approx 4.2 \times 10^7 /s$ . 

The wavelength of parametric fluorescence generated here ranges from $\approx 430$ nm to above 1000 nm. The collection efficiency and spectral response of the optical systems could vary more than an order of magnitude in such a broad wavelength range. Therefore we consider the degenerate parametric fluorescence at $\lambda=810$ nm to compare the experimentally determined fluorescence photon flux with the theoretically integrated photon flux (see Eq.~(\ref{eq:int_flux})). 

\begin{figure}[!tbp]
\includegraphics[width=1.0 \columnwidth]{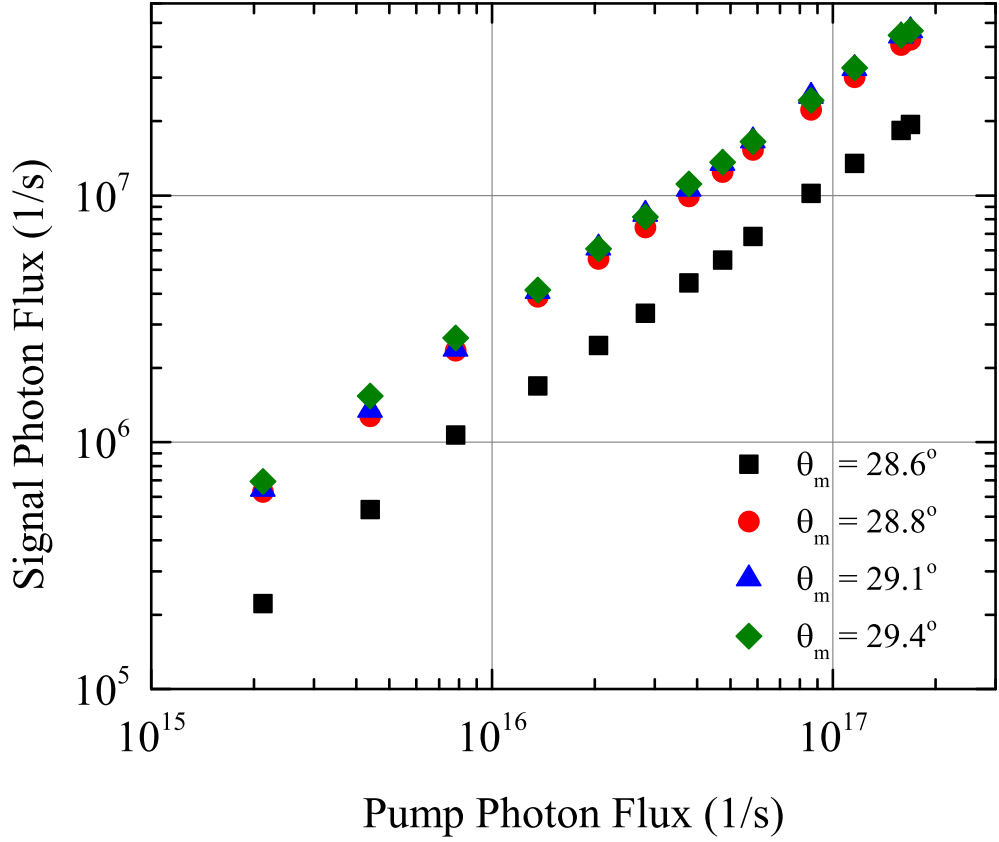}
\caption{{\bf Parametric fluorescence flux at 810nm.} Integrated degenerate parametric fluorescence flux at $\lambda_s=810$ nm as a function of the incident pump flux for the collinear case and three phase-matching angles $\theta_{m}$ of the angle-resolved images shown in Fig.~\ref{fig:ARImages}. Signal flux is linearly proportional to the pump flux over two order of magnitude, confirming that the dominant signal is \emph{spontaneous} parametric fluorescence as described by Eq. (\ref{eq:int_flux}). The slopes $\eta=N_s / N_p$ are $1.16 \times 10^{-10}$ for $\theta_m = 28.6^{\circ}$, $ 2.6 \times 10^{-10}$ for $\theta_m = 28.8^{\circ}$, and $2.8 \times 10^{-10}$  for $\theta_m = 29.1^{\circ}$ and $29.4^{\circ}$. \label{fig:PowerDep}}
\end{figure}

The integrated degenerate parametric fluorescence flux at $\lambda_s=810$ nm as a function of the incident pump flux are shown in Fig.~\ref{fig:PowerDep}. First, we determine the overall collection efficiency and the spectral response of the imaging and spectroscopy system by passing a laser beam of $\lambda=810$ nm with a known photon flux through the same optical path as the parametric fluorescence. We then integrate the signal in angle-resolved images as shown in Fig. \ref{fig:ARImages}. Taking into account the system response at $\lambda=810$ nm, we can determine the absolute fluorescence flux experimentally within roughly 20\% error. The relative fluorescence fluxes, determined by the linearity of the CCD camera, is within 1\% error. The absolute value of pump flux is within roughly 10\% error. The relative pump fluxes, as varied by a combination of a half-wave plate and a polarizer and limited by the linearity of the power meter, are known within a few percent. Thus we can investigate the pump flux (power) dependence of parametric fluorescence flux with precision. Fluorescence signal flux is linearly proportional to the pump flux over two order of magnitude, confirming that the dominant signal is \emph{spontaneous} parametric fluorescence as described by Eq. (\ref{eq:int_flux}). The slopes correspond to the efficiency coefficients. Considering both signal and idler fluorescence, we determined $\eta=2 \eta_s$ to be approximately $2.8 \times 10^{-10}$. Experimental and theoretical values of $\eta=2 \eta_s$ for selected phase-matching angles are listed in Table~\ref{tab:PhotonFlux}.

\subsection{Instrument Spectral Response Function}
Parametric fluorescence spectra can also be used to calibrate the instrument spectral response function (ISRF) of the imaging spectroscopy system. According to Eq.~(\ref{eq:int_flux}), which is valid for non-collinear cases, we can deduce a parameter $S \equiv N_s \times \lambda_s^4 \, \lambda_i^2=2 \, (2 \pi)^4 \, \hbar \, c \, {d_{eff}}^2 \, L \, N_p \, d \lambda_s \,/(\epsilon_0 \, {n_p}^2 \,) $ \cite{Koch1995}. $S$ is a wavelength-independent constant for a given pump wavelength and geometry. We define a generalized spectral function, $S(\lambda_s) \equiv N_s(\lambda_s) \, \lambda_s^4 \, \lambda_i^2$, for both calculated and experimental angle-resolved spectra. The calculated $S_{sim}(\lambda_s)$ exhibit less than 1\% variation between  $\lambda=460$ and 1000 nm. Experimentally, $S^*_{exp}(\lambda_s)=N^*_s(\lambda)\,  \lambda_s^4 \, \lambda_i^2$ can be determined from the integration of an angle-resolved spectrum $N^*_s(\lambda, \theta_s)$ over $\theta_s$. $S_{exp}(\lambda)$ represents the relative ISRF of the imaging spectroscopy system, including the optical components, grating, and CCD camera along the fluorescence collection optical path. The value of the ISRF at a fixed wavelength can then be used to determine the absolute ISRF across the parametric fluorescence wavelength range. The effective efficiency, defined as (\# of photo-generated electrons / \# of photons), is $\approx$20\% at $\lambda=810$ nm in our experiments. $S^*_{exp}(\lambda)$ and $S_{sim}(\lambda)$ for phase-matching angles $\theta_m = 29.1^\circ$ and $29.4^\circ$ are shown in Fig.~\ref{fig:PFeffRatio}. The stray scattered pump laser signal becomes increasingly difficult to subtract from these angle-resolved spectra, leading to a distorted $S^*_{exp}(\lambda)$.  We thus use $S^*_{exp}(\lambda)$ at $\theta_m = 29.1^\circ$ to deduce the ISRF of our optical setup shown in Fig.~\ref{fig:PFExpSetup}.


\begin{acknowledgments}
We thank Brage Golding and John A. McGuire for discussions. This work was supported by grant DMR-0955944 from the National Science Foundation and by internal Strategic Initiative Projects of the College of Natural Science at Michigan State University.
\end{acknowledgments}

\section{Appendix:  Calculation of Fluorescence Flux}
Here we describe the integration over $\vect{\xi}$ of Eq.~(\ref{eq:photon_flux}) for the calculation of the angular distribution of fluorescence flux here. The mismatch wave vector $\Delta \vect{k}=\vect{k}'_s+\vect{k}'_i-\vect{k}'_p$ can be decomposed into longitudinal ($\Delta k_z \,\hat{z}$) and transverse ($\vect{\kappa}$) parts: 
\begin{align}
\Delta k_z &= k'_{sz}+k'_{iz} - k'_{pz} \nonumber \\
 &= \sqrt{{k'_s}^2 - \kappa _s^2} + \sqrt{ {k'_i}^2- \kappa _i^2} - k'_{pz} \, , \; \rm{and}  \label{eq:z_momentum} \nonumber \\
\vect{\kappa} &= \vect{\kappa}_s + \vect{\kappa}_i \,. \nonumber
\end{align}

Here $k'_s = {n_s} {\omega _s} / c$, $k'_i = {n_i} {\omega _i} / c$, and $k'_p = {n_p} {\omega _p} / c$ are the wave numbers for the signal, idler, and pump, respectively. $\vect{\kappa}_s$ ($\vect{\kappa_i}$) is the transverse wave vector of the signal (idler) wave. The phase-matching function $\Phi$ in Eq.~(\ref{eq:Phi}) is a function of three variables: $\omega_s$, $\vect{\kappa}_s$ and $\vect{\kappa}_i$. Considering energy conservation $\omega_p = \omega_s + \omega_i$, we can carry out the integration over $\vect{\xi}$ for two independent variables, $\omega_s$ and $\vect{\kappa}_s$. Using $\vect{\xi} = \vect{\xi}_s + \vect{\xi}_i = (\vect{\kappa_s}+\vect{\kappa_i)} \, W $, we rewrite Eq.~(\ref{eq:photon_flux}) as

\begin{widetext}
\begin{equation} 
N_s(\omega _s, \vect{\kappa}_s) = \frac{\hbar \, {d_{eff}}^2 \, {\omega _s} \, {\omega _i} \, {\omega_p} \, L^2 \, N_p }{8 \pi^4 \, c^3 \, \epsilon _0 \, n_s \, n_i \, n_p} \, d\omega _s \, d^2 \vect{\kappa}_s \int d^2 \vect{\xi}_i \, \exp \left( - \frac{1}{2} |\vect{\xi}_s+\vect{\xi}_i|^2 \right) \cdot \rm{sinc}^2 \left( {\frac{1}{2}\,L\,\Delta {k_z}} \right) \nonumber
\end{equation}
\end{widetext}

We further simplify the numerical integration by (a) considering that the $\rm{sinc}^2$ term is a constant for a given set of $\omega_s$ and $\vect{\kappa}_s$, and (b) applying a saddle-point approximation. 

$\vect{\kappa}_s$ and $\vect{\kappa}_i$ form an angle $\varphi$ in the $x-y$ plane of the crystal, where $\varphi = 0$ in the anti-parallel case. Under a pump wave vector along $\hat{z}$, the phase-matching condition is met mostly for $\vect{k}'_s+\vect{k}'_i=0$; i.e., $\varphi \approx 0$.  The Gaussian term of the integrand can thus be separated and integrated with a saddle-point approximation for $\varphi \approx 0$ and $\xi_s\approx \xi_i$:

\begin{widetext}
\begin{align} 
\int d^2 \vect{\xi}_i \, \exp \left( - \frac{1}{2} (\xi_s^2+\xi_i^2-2 \, \xi_s \,\xi_i \, \cos(\varphi)) \right) 
&\approx  \int \xi_i d \xi_i \, \exp \left( - \frac{1}{2} \left(\xi_s - \xi_i \right)^2 \right)\int d \varphi \exp \left(-\frac{1}{2} \xi_s \, \xi_i \, \varphi^2 \right ) \nonumber \\
&\approx  \sqrt{2 \pi} \int d\xi_i \exp \left(-\frac{1}{2} \left(\xi_s - \xi_i \right )^2 \right ) \nonumber \, . \label{eq:saddle-point}
\end{align}
\end{widetext}

The integrand is a maximum at $\xi_s=\xi_i$, where pairs of signal and idler photons are emitted in approximately opposite conical directions. Applying the above two approximations, we obtain

\begin{widetext}
\begin{align}
N_s(\omega _s, \kappa_s, \phi) = \frac{\hbar \, {d_{eff}}^2 \, {\omega _s} \, {\omega _i} \, {\omega_p} \, L^2 \, N_p }{8 \pi^4 \, c^3 \, \epsilon _0 \, n_s \, n_i \, n_p} \sqrt{2\pi} \, d\omega _s \, \kappa_s \, d \kappa_s \, d\phi \int d \xi_i \,  e^{- \frac{1}{2} (\xi_s-\xi_i)^2} \cdot \rm{sinc}^2 \left( {\frac{1}{2}\,L\,\Delta {k_z}} \right), \nonumber
\end{align}
where $d^2 \vect{\kappa}_s = \kappa_s \, d \kappa_s \, d \phi$, $\phi$ is the azimuthal angle. 
\end{widetext}

$N_s(\omega _s, \kappa_s, \phi)$ is isotropic in $\phi$. It can be expressed as $N_s(\lambda _s, \theta_s)$, a function of experimentally measurable signal angle $\theta_s$ and wavelength $\lambda_s$ by considering that $\kappa_{s}=k_{s} \sin \theta_{s}$ and $\omega_s={2 \pi c}/\lambda_s$: 

\begin{widetext}
\begin{equation} 
N_s(\lambda _s, \theta_s) = \frac{2\pi\, \sqrt{2\pi} \, \hbar \,  {d_{eff}}^2 \, {\omega_p} \, L^2 \, N_p }{ \epsilon _0 \, n_s \, n_i \, n_p \, {\lambda_s^5}\, {\lambda_i}}   \, \sin (2\theta_s) \, d\lambda_s  \, d \theta_s \int d \xi_i \, e^{-\frac{1}{2} (\xi_s-\xi_i)^2} \cdot \rm{sinc}^2 \left( {\frac{1}{2}\,L\,\Delta {k_z}} \right). \label{eq:photon_flux_sim}
\end{equation}
\end{widetext}

\begin{figure}[!tbp]
\includegraphics[width=1.0 \columnwidth]{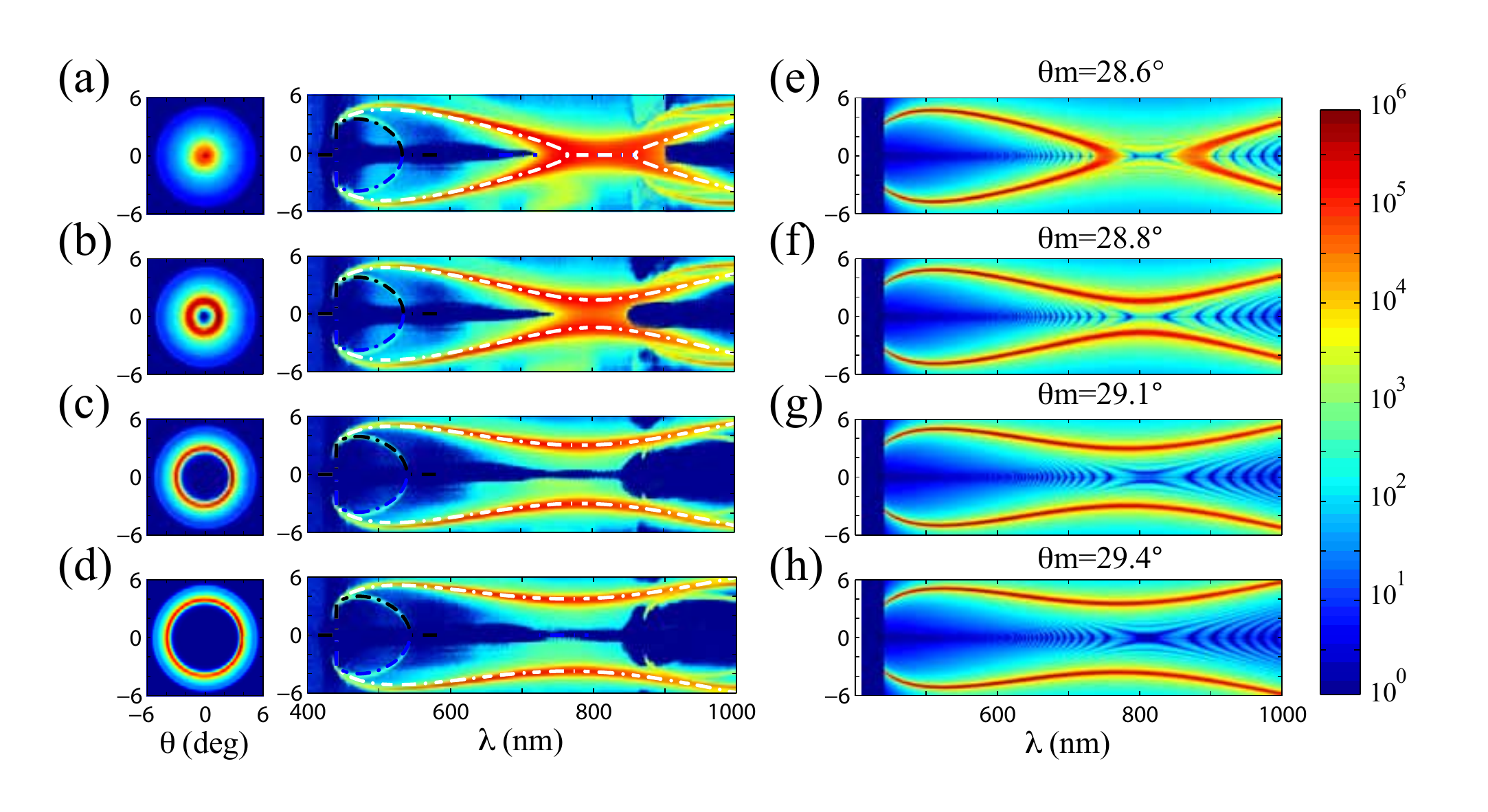}
\caption{{\bf Angle-resolved spectra of parametric fluorescence.} (a)-(d) Experimental angle-resolved parametric fluorescence images (left panel) and spectra (right panel) for a phase-matching angle $\theta_{m}=28.6^\circ, 28.8^\circ, 29.1^\circ,$ and $29.4^\circ$, respectively. (e)-(h) Theoretical fluorescence flux calculated according to Eq.~(\ref{eq:photon_flux}) (see also Appendix). The tuning curves for the perfect phase-matching condition are indicated by the white dashed lines on experimental imaging spectra. The color palette represents the calculated photon flux with $\Delta \lambda_s=1$ nm and $\Delta \theta_s=0.5$ mrad on a logarithmic scale. \label{fig:ARSpectra}}
\end{figure}

\begin{figure}[!tbp]
\includegraphics[width=1.0 \columnwidth]{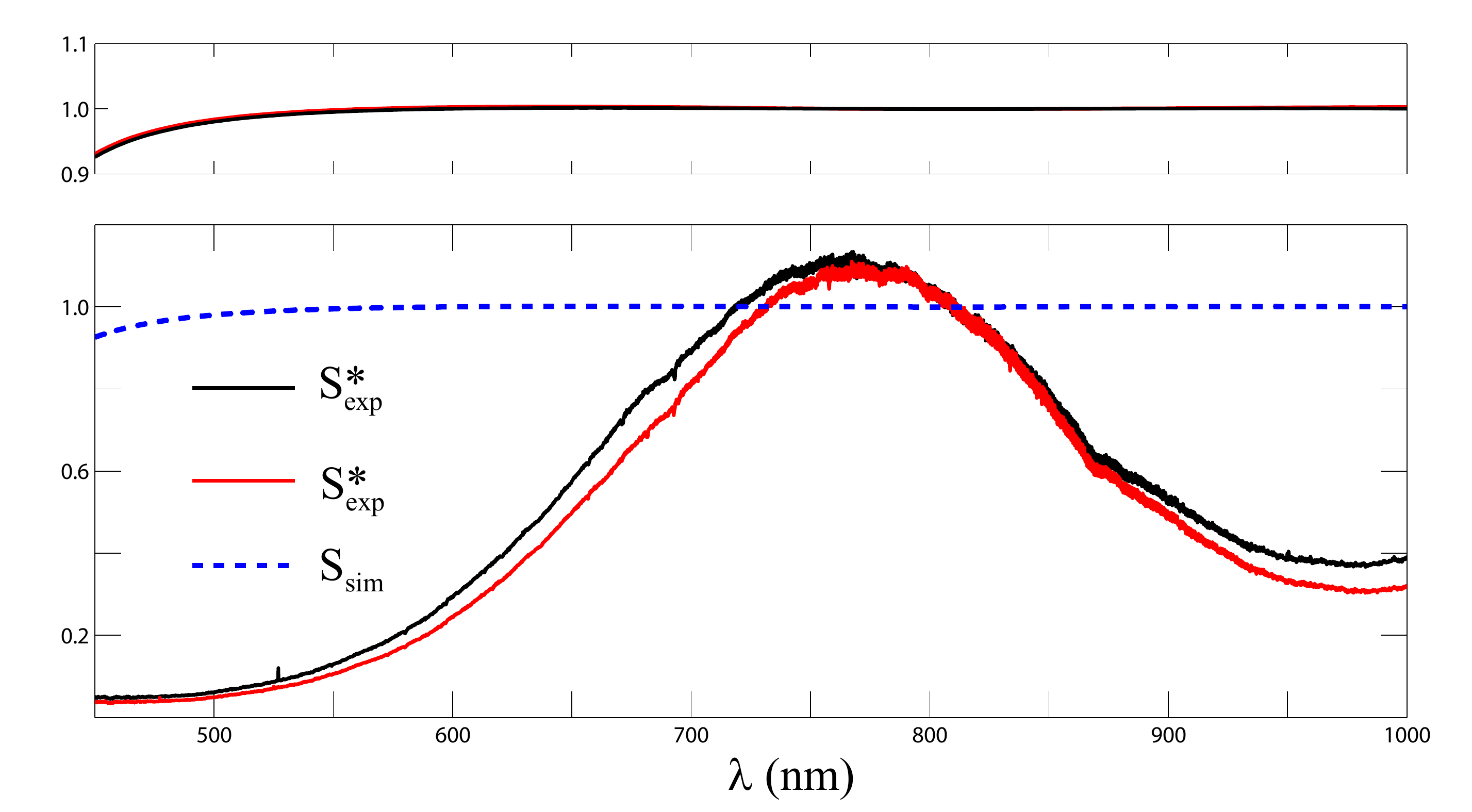}
\caption{{\bf Instrument spectral response function.} The normalized spectral density function $S(\lambda)$ for  $\theta_m=29.1^\circ$ and $29.4^\circ$. $S_{sim}$ is determined from the integration over $\theta_s=-7.5$ to $7.5^\circ$ of the calculated angle-resolved spectra as shown in Fig.~\ref{fig:ARSpectra} (Eq.~(\ref{eq:photon_flux})). The black and red curves are $S_{sim}$ or $S^*_{exp}$ for $\theta_m=29.4^\circ$ and $29.1^\circ$, respectively. The value of $S_{sim}$ is a constant with 1\% standard deviation across wavelengths from 500 to 1000 nm, validating the calculated angular fluorescence spectra and Eq.~ (\ref{eq:int_flux}). $S^*_{exp}(\lambda)=N^*_s(\lambda) \, {\lambda_s}^4 \, {\lambda_i}^2$, where $N^*_s(\lambda)$ is the integration over $\theta_s\approx-15$ to $15^\circ$ of the experimental imaging spectra $N^*_s(\lambda_s, \theta_s)$. $S^*_{exp}(\lambda)$ represents a relative instrument spectral response function (ISRF) of the optical spectroscopy system, including optical filters and a Glan-Thompson polarizer along the path of the fluorescence , a liquid-nitrogen-cooled CCD (PI-Acton Spec-10:400BR), and a 300g/mm plane ruled reflectance grating with 1000-nm blaze wavelength (PI-Acton 750-1-030-1). The polarization of the fluorescence is vertically polarized. The absolution ISRF (or collection efficiency) can be deduced by calibrating $S^*_{exp}(\lambda)$ at a fixed wavelength such as $\lambda=810$ nm in our experiments.  
\label{fig:PFeffRatio}} 
\end{figure}

The equation for $N_s(\lambda_s, \theta_s)$ above, together with $\omega_p=\omega_s+\omega_i$ and $\textbf{k}_{p} = \textbf{k}_{s}+ \textbf{k}_{i}$, is used in the numerical integration to obtain the theoretical angular distribution of parametric fluorescence shown in Fig.~\ref{fig:ARSpectra}. 

\bibliography{PF}

\begin{thebibliography}{25}
\expandafter\ifx\csname natexlab\endcsname\relax\def\natexlab#1{#1}\fi
\expandafter\ifx\csname bibnamefont\endcsname\relax
  \def\bibnamefont#1{#1}\fi
\expandafter\ifx\csname bibfnamefont\endcsname\relax
  \def\bibfnamefont#1{#1}\fi
\expandafter\ifx\csname citenamefont\endcsname\relax
  \def\citenamefont#1{#1}\fi
\expandafter\ifx\csname url\endcsname\relax
  \def\url#1{\texttt{#1}}\fi
\expandafter\ifx\csname urlprefix\endcsname\relax\def\urlprefix{URL }\fi
\providecommand{\bibinfo}[2]{#2}
\providecommand{\eprint}[2][]{\url{#2}}

\bibitem[{\citenamefont{Shih and Alley}(1988)}]{Shih1988}
\bibinfo{author}{\bibfnamefont{Y.~H.} \bibnamefont{Shih}} \bibnamefont{and}
  \bibinfo{author}{\bibfnamefont{C.~O.} \bibnamefont{Alley}},
  \bibinfo{journal}{Phys. Rev. Lett.} \textbf{\bibinfo{volume}{61}},
  \bibinfo{pages}{2921} (\bibinfo{year}{1988}),
  \urlprefix\url{http://link.aps.org/doi/10.1103/PhysRevLett.61.2921}.

\bibitem[{\citenamefont{Shih et~al.}(1994)\citenamefont{Shih, Sergienko, Rubin,
  Kiess, and Alley}}]{Shih1994}
\bibinfo{author}{\bibfnamefont{Y.~H.} \bibnamefont{Shih}},
  \bibinfo{author}{\bibfnamefont{A.~V.} \bibnamefont{Sergienko}},
  \bibinfo{author}{\bibfnamefont{M.~H.} \bibnamefont{Rubin}},
  \bibinfo{author}{\bibfnamefont{T.~E.} \bibnamefont{Kiess}}, \bibnamefont{and}
  \bibinfo{author}{\bibfnamefont{C.~O.} \bibnamefont{Alley}},
  \bibinfo{journal}{Phys. Rev. A} \textbf{\bibinfo{volume}{50}},
  \bibinfo{pages}{23} (\bibinfo{year}{1994}),
  \urlprefix\url{http://link.aps.org/doi/10.1103/PhysRevA.50.23}.

\bibitem[{\citenamefont{Kwiat et~al.}(1995)\citenamefont{Kwiat, Mattle,
  Weinfurter, Zeilinger, Sergienko, and Shih}}]{Kwiat1995}
\bibinfo{author}{\bibfnamefont{P.~G.} \bibnamefont{Kwiat}},
  \bibinfo{author}{\bibfnamefont{K.}~\bibnamefont{Mattle}},
  \bibinfo{author}{\bibfnamefont{H.}~\bibnamefont{Weinfurter}},
  \bibinfo{author}{\bibfnamefont{A.}~\bibnamefont{Zeilinger}},
  \bibinfo{author}{\bibfnamefont{A.~V.} \bibnamefont{Sergienko}},
  \bibnamefont{and} \bibinfo{author}{\bibfnamefont{Y.}~\bibnamefont{Shih}},
  \bibinfo{journal}{Phys. Rev. Lett.} \textbf{\bibinfo{volume}{75}},
  \bibinfo{pages}{4337} (\bibinfo{year}{1995}),
  \urlprefix\url{http://link.aps.org/doi/10.1103/PhysRevLett.75.4337}.

\bibitem[{\citenamefont{Shih and Shih}(2003)}]{Shih2003}
\bibinfo{author}{\bibfnamefont{Y.}~\bibnamefont{Shih}} \bibnamefont{and}
  \bibinfo{author}{\bibfnamefont{Y.}~\bibnamefont{Shih}},
  \bibinfo{journal}{Reports on Progress in Physics}
  \textbf{\bibinfo{volume}{66}}, \bibinfo{pages}{1009} (\bibinfo{year}{2003}),
  ISSN \bibinfo{issn}{0034-4885},
  \urlprefix\url{http://dx.doi.org/10.1088/0034-4885/66/6/203}.

\bibitem[{\citenamefont{Louisell et~al.}(1961)\citenamefont{Louisell, Yariv,
  and Siegman}}]{Louisell1961}
\bibinfo{author}{\bibfnamefont{W.~H.} \bibnamefont{Louisell}},
  \bibinfo{author}{\bibfnamefont{A.}~\bibnamefont{Yariv}}, \bibnamefont{and}
  \bibinfo{author}{\bibfnamefont{A.~E.} \bibnamefont{Siegman}},
  \bibinfo{journal}{Phys. Rev.} \textbf{\bibinfo{volume}{124}},
  \bibinfo{pages}{1646} (\bibinfo{year}{1961}),
  \urlprefix\url{http://link.aps.org/doi/10.1103/PhysRev.124.1646}.

\bibitem[{\citenamefont{Gordon et~al.}(1963)\citenamefont{Gordon, Louisell, and
  Walker}}]{Gordon1963}
\bibinfo{author}{\bibfnamefont{J.~P.} \bibnamefont{Gordon}},
  \bibinfo{author}{\bibfnamefont{W.~H.} \bibnamefont{Louisell}},
  \bibnamefont{and} \bibinfo{author}{\bibfnamefont{L.~R.}
  \bibnamefont{Walker}}, \bibinfo{journal}{Phys. Rev.}
  \textbf{\bibinfo{volume}{129}}, \bibinfo{pages}{481} (\bibinfo{year}{1963}),
  \urlprefix\url{http://link.aps.org/doi/10.1103/PhysRev.129.481}.

\bibitem[{\citenamefont{Kleinman}(1968)}]{Kleinman1968}
\bibinfo{author}{\bibfnamefont{D.~A.} \bibnamefont{Kleinman}},
  \bibinfo{journal}{Phys. Rev. A} \textbf{\bibinfo{volume}{174}},
  \bibinfo{pages}{1027} (\bibinfo{year}{1968}),
  \urlprefix\url{http://link.aps.org/doi/10.1103/PhysRev.174.1027}.

\bibitem[{\citenamefont{Harris et~al.}(1967)\citenamefont{Harris, Oshman, and
  Byer}}]{Harris1967}
\bibinfo{author}{\bibfnamefont{S.~E.} \bibnamefont{Harris}},
  \bibinfo{author}{\bibfnamefont{M.~K.} \bibnamefont{Oshman}},
  \bibnamefont{and} \bibinfo{author}{\bibfnamefont{R.~L.} \bibnamefont{Byer}},
  \bibinfo{journal}{Phys. Rev. Lett.} \textbf{\bibinfo{volume}{18}},
  \bibinfo{pages}{732} (\bibinfo{year}{1967}),
  \urlprefix\url{http://link.aps.org/doi/10.1103/PhysRevLett.18.732}.

\bibitem[{\citenamefont{Giallorenzi and Tang}(1968)}]{Giallorenzi1968}
\bibinfo{author}{\bibfnamefont{T.~G.} \bibnamefont{Giallorenzi}}
  \bibnamefont{and} \bibinfo{author}{\bibfnamefont{C.~L.} \bibnamefont{Tang}},
  \bibinfo{journal}{Phys. Rev.} \textbf{\bibinfo{volume}{166}},
  \bibinfo{pages}{225} (\bibinfo{year}{1968}),
  \urlprefix\url{http://link.aps.org/doi/10.1103/PhysRev.166.225}.

\bibitem[{\citenamefont{Byer and Harris}(1968)}]{Byer1968}
\bibinfo{author}{\bibfnamefont{R.~L.} \bibnamefont{Byer}} \bibnamefont{and}
  \bibinfo{author}{\bibfnamefont{S.~E.} \bibnamefont{Harris}},
  \bibinfo{journal}{Phys. Rev.} \textbf{\bibinfo{volume}{168}},
  \bibinfo{pages}{1064} (\bibinfo{year}{1968}),
  \urlprefix\url{http://link.aps.org/doi/10.1103/PhysRev.168.1064}.

\bibitem[{\citenamefont{Choy and Byer}(1976)}]{Choy1976}
\bibinfo{author}{\bibfnamefont{M.~M.} \bibnamefont{Choy}} \bibnamefont{and}
  \bibinfo{author}{\bibfnamefont{R.~L.} \bibnamefont{Byer}},
  \bibinfo{journal}{Phys. Rev. B} \textbf{\bibinfo{volume}{14}},
  \bibinfo{pages}{1693} (\bibinfo{year}{1976}),
  \urlprefix\url{http://link.aps.org/doi/10.1103/PhysRevB.14.1693}.

\bibitem[{\citenamefont{Shoji et~al.}(1997)\citenamefont{Shoji, Kondo,
  Kitamoto, Shirane, and Ito}}]{Shoji1997}
\bibinfo{author}{\bibfnamefont{I.}~\bibnamefont{Shoji}},
  \bibinfo{author}{\bibfnamefont{T.}~\bibnamefont{Kondo}},
  \bibinfo{author}{\bibfnamefont{A.}~\bibnamefont{Kitamoto}},
  \bibinfo{author}{\bibfnamefont{M.}~\bibnamefont{Shirane}}, \bibnamefont{and}
  \bibinfo{author}{\bibfnamefont{R.}~\bibnamefont{Ito}}, \bibinfo{journal}{JOSA
  B} \textbf{\bibinfo{volume}{14}}, \bibinfo{pages}{2268}
  (\bibinfo{year}{1997}),
  \urlprefix\url{http://dx.doi.org/10.1364/JOSAB.14.002268}.

\bibitem[{\citenamefont{Shoji et~al.}(2002)\citenamefont{Shoji, Kondo, and
  Ito}}]{Shoji2002}
\bibinfo{author}{\bibfnamefont{I.}~\bibnamefont{Shoji}},
  \bibinfo{author}{\bibfnamefont{T.}~\bibnamefont{Kondo}}, \bibnamefont{and}
  \bibinfo{author}{\bibfnamefont{R.}~\bibnamefont{Ito}},
  \bibinfo{journal}{Optical and Quantum Electronics}
  \textbf{\bibinfo{volume}{34}}, \bibinfo{pages}{797} (\bibinfo{year}{2002}),
  \urlprefix\url{http://dx.doi.org/10.1023/A:1016545417478}.

\bibitem[{\citenamefont{Zeldovich and Klyshko}(1969)}]{Zeldovich1969}
\bibinfo{author}{\bibfnamefont{B.~Y.} \bibnamefont{Zeldovich}}
  \bibnamefont{and} \bibinfo{author}{\bibfnamefont{D.~N.}
  \bibnamefont{Klyshko}}, \bibinfo{journal}{JETP Letters}
  \textbf{\bibinfo{volume}{9}}, \bibinfo{pages}{40} (\bibinfo{year}{1969}),
  \urlprefix\url{http://www.jetpletters.ac.ru/ps/1639/article_25275.shtml}.

\bibitem[{\citenamefont{Burnham and Weinberg}(1970)}]{Burnham1970}
\bibinfo{author}{\bibfnamefont{D.~C.} \bibnamefont{Burnham}} \bibnamefont{and}
  \bibinfo{author}{\bibfnamefont{D.~L.} \bibnamefont{Weinberg}},
  \bibinfo{journal}{Phys. Rev. Lett.} \textbf{\bibinfo{volume}{25}},
  \bibinfo{pages}{84} (\bibinfo{year}{1970}),
  \urlprefix\url{http://link.aps.org/doi/10.1103/PhysRevLett.25.84}.

\bibitem[{\citenamefont{Chen et~al.}(1985)\citenamefont{Chen, Wu, Jiang, and
  You}}]{Chen1985}
\bibinfo{author}{\bibfnamefont{C.}~\bibnamefont{Chen}},
  \bibinfo{author}{\bibfnamefont{B.}~\bibnamefont{Wu}},
  \bibinfo{author}{\bibfnamefont{A.}~\bibnamefont{Jiang}}, \bibnamefont{and}
  \bibinfo{author}{\bibfnamefont{G.}~\bibnamefont{You}}, \bibinfo{journal}{Sci.
  Sin. Ser. B} \textbf{\bibinfo{volume}{28}}, \bibinfo{pages}{235}
  (\bibinfo{year}{1985}).

\bibitem[{\citenamefont{Boeuf et~al.}(2000)\citenamefont{Boeuf, Branning,
  Chaperot, Dauler, Guerin, Jaeger, Muller, and Migdall}}]{Boeuf2000}
\bibinfo{author}{\bibfnamefont{N.}~\bibnamefont{Boeuf}},
  \bibinfo{author}{\bibfnamefont{D.}~\bibnamefont{Branning}},
  \bibinfo{author}{\bibfnamefont{I.}~\bibnamefont{Chaperot}},
  \bibinfo{author}{\bibfnamefont{E.}~\bibnamefont{Dauler}},
  \bibinfo{author}{\bibfnamefont{S.}~\bibnamefont{Guerin}},
  \bibinfo{author}{\bibfnamefont{G.}~\bibnamefont{Jaeger}},
  \bibinfo{author}{\bibfnamefont{A.}~\bibnamefont{Muller}}, \bibnamefont{and}
  \bibinfo{author}{\bibfnamefont{A.}~\bibnamefont{Migdall}},
  \bibinfo{journal}{Opt. Eng.} \textbf{\bibinfo{volume}{39}},
  \bibinfo{pages}{1016} (\bibinfo{year}{2000}), ISSN \bibinfo{issn}{00913286},
  \urlprefix\url{http://dx.doi.org/10.1117/1.602464}.

\bibitem[{\citenamefont{Koch et~al.}(1995)\citenamefont{Koch, Cheung, Moore,
  Chakmakjian, and Liu}}]{Koch1995}
\bibinfo{author}{\bibfnamefont{K.}~\bibnamefont{Koch}},
  \bibinfo{author}{\bibfnamefont{E.~C.} \bibnamefont{Cheung}},
  \bibinfo{author}{\bibfnamefont{G.~T.} \bibnamefont{Moore}},
  \bibinfo{author}{\bibfnamefont{S.~H.} \bibnamefont{Chakmakjian}},
  \bibnamefont{and} \bibinfo{author}{\bibfnamefont{J.~M.} \bibnamefont{Liu}},
  \bibinfo{journal}{IEEE Journal of Quantum Electronics}
  \textbf{\bibinfo{volume}{31}}, \bibinfo{pages}{769} (\bibinfo{year}{1995}),
  ISSN \bibinfo{issn}{00189197},
  \urlprefix\url{http://dx.doi.org/10.1109/3.375922}.

\bibitem[{\citenamefont{Hong and Mandel}(1985)}]{Hong1985}
\bibinfo{author}{\bibfnamefont{C.~K.} \bibnamefont{Hong}} \bibnamefont{and}
  \bibinfo{author}{\bibfnamefont{L.}~\bibnamefont{Mandel}},
  \bibinfo{journal}{Physical Review A} \textbf{\bibinfo{volume}{31}},
  \bibinfo{pages}{2409} (\bibinfo{year}{1985}),
  \urlprefix\url{http://link.aps.org/doi/10.1103/PhysRevA.31.2409}.

\bibitem[{\citenamefont{Ling et~al.}(2008)\citenamefont{Ling, Lamas-Linares,
  and Kurtsiefer}}]{Ling2008}
\bibinfo{author}{\bibfnamefont{A.}~\bibnamefont{Ling}},
  \bibinfo{author}{\bibnamefont{Lamas-Linares}}, \bibnamefont{and}
  \bibinfo{author}{\bibfnamefont{C.}~\bibnamefont{Kurtsiefer}},
  \bibinfo{journal}{Phys. Rev. A} \textbf{\bibinfo{volume}{77}},
  \bibinfo{pages}{043834} (\bibinfo{year}{2008}),
  \urlprefix\url{http://link.aps.org/doi/10.1103/PhysRevA.77.043834}.

\bibitem[{\citenamefont{Mitchell}(2009)}]{Mitchell2009}
\bibinfo{author}{\bibfnamefont{M.~W.} \bibnamefont{Mitchell}},
  \bibinfo{journal}{Phys. Rev. A} \textbf{\bibinfo{volume}{79}},
  \bibinfo{pages}{043835} (\bibinfo{year}{2009}),
  \urlprefix\url{http://link.aps.org/doi/10.1103/PhysRevA.79.043835}.

\bibitem[{\citenamefont{Rubin}(1996)}]{Rubin1996}
\bibinfo{author}{\bibfnamefont{M.~H.} \bibnamefont{Rubin}},
  \bibinfo{journal}{Phys. Rev. A} \textbf{\bibinfo{volume}{54}},
  \bibinfo{pages}{5349} (\bibinfo{year}{1996}),
  \urlprefix\url{http://link.aps.org/doi/10.1103/PhysRevA.54.5349}.

\bibitem[{\citenamefont{Fedorov et~al.}(2008)\citenamefont{Fedorov, Efremov,
  Volkov, Moreva, Straupe, and Kulik}}]{Fedorov2008}
\bibinfo{author}{\bibfnamefont{M.}~\bibnamefont{Fedorov}},
  \bibinfo{author}{\bibfnamefont{M.}~\bibnamefont{Efremov}},
  \bibinfo{author}{\bibfnamefont{P.}~\bibnamefont{Volkov}},
  \bibinfo{author}{\bibfnamefont{E.}~\bibnamefont{Moreva}},
  \bibinfo{author}{\bibfnamefont{S.}~\bibnamefont{Straupe}}, \bibnamefont{and}
  \bibinfo{author}{\bibfnamefont{S.}~\bibnamefont{Kulik}},
  \bibinfo{journal}{Physical Review A} \textbf{\bibinfo{volume}{77}},
  \bibinfo{pages}{1} (\bibinfo{year}{2008}), ISSN \bibinfo{issn}{1050-2947},
  \urlprefix\url{http://link.aps.org/doi/10.1103/PhysRevA.77.032336}.

\bibitem[{\citenamefont{Klein et~al.}(2003)\citenamefont{Klein, Kugel,
  Maillard, Sifi, and Polg\'{a}r}}]{Klein2003}
\bibinfo{author}{\bibfnamefont{R.~S.} \bibnamefont{Klein}},
  \bibinfo{author}{\bibfnamefont{G.~E.} \bibnamefont{Kugel}},
  \bibinfo{author}{\bibfnamefont{A.}~\bibnamefont{Maillard}},
  \bibinfo{author}{\bibfnamefont{A.}~\bibnamefont{Sifi}}, \bibnamefont{and}
  \bibinfo{author}{\bibfnamefont{K.}~\bibnamefont{Polg\'{a}r}},
  \bibinfo{journal}{Optical Materials} \textbf{\bibinfo{volume}{22}},
  \bibinfo{pages}{163 } (\bibinfo{year}{2003}), ISSN \bibinfo{issn}{0925-3467},
  \urlprefix\url{http://dx.doi.org/10.1016/S0925-3467(02)00360-9}.

\bibitem[{\citenamefont{Migdall}(2000)}]{Migdall2000}
\bibinfo{author}{\bibfnamefont{A.}~\bibnamefont{Migdall}},
  \bibinfo{journal}{Optical Engineering} \textbf{\bibinfo{volume}{39}},
  \bibinfo{pages}{1016} (\bibinfo{year}{2000}), ISSN \bibinfo{issn}{00913286},
  \urlprefix\url{http://dx.doi.org/10.1117/1.602464}.

\end{thebibliography}

\pagebreak

\begin{table} [!tbp]
\caption{Parametric fluorescence efficiency $\eta = 2 N_s/N_p$ for $810\pm 0.5$ nm}
    \begin{tabular}{|c|c|c|c|c|}
        \hline
  $\theta_m$ & Exp. & Sim. [Eq. (\ref{eq:photon_flux_sim})] & Th. [Eq.~(\ref{eq:int_flux})] & $n_p$  \\ \hline
 $ \, 28.6^{\circ}$ & $1.3 \times 10^{-10}$  & $2.5 \times 10^{-11} $ & $--$& 1.6612 \\ \hhline{-----} 
 $ \, 28.8^{\circ}$ & $2.8 \times 10^{-10}$  & $2.5 \times 10^{-10}$ & \multirow{3}{*}{$2.6 \times 10^{-10}$} & 1.6608 \\ \hhline{---~-}
 $ \, 29.1^{\circ}$ & $2.7 \times 10^{-10}$  & \multirow{2}{*}{$2.6 \times 10^{-10}$}  &  & 1.6602 \\ \hhline{--~~-}
 $ \, 29.4^{\circ}$ & $2.7 \times 10^{-10} $ &   &  & 1.6599\\
        \hline
    \end{tabular} \label{tab:PhotonFlux}
\end{table}

\end{document}